\documentclass[aps,pra,preprint,showpacs]{revtex4}
\usepackage{graphicx}
\usepackage{amsmath,amssymb,amsfonts}
\usepackage{natbib}

\begin{document}

\title{Aspects of quantum coherence in the optical Bloch equations}

\author{A. S. Sanz}
 \email[Present Address: Instituto de Matem\'aticas y F\'{\i}sica
 Fundamental, CSIC, Serrano 123, 28006 Madrid, Spain. \\
 E--mail address: ]{cees374@imaff.cfmac.csic.es}

\author{H. Han}
 \email[Present Address: Department of Chemistry,
  Queen's University, Kingston ON, Canada K7L 3N6. \\
  E--mail: ]{hanh@chem.queensu.ca}

\author{P. Brumer}
 \email[E--mail address: ]{pbrumer@chem.utoronto.ca}
 \affiliation{Chemical Physics Theory Group,
  Department of Chemistry, \\
and
  Center for Quantum Information and Quantum Control, \\
University of Toronto,
  Toronto Ontario, Canada M5S 3H6.}

\date{\today}

\begin{abstract}
Aspects of coherence and decoherence are analyzed within the optical
Bloch equations.
By rewriting the analytic solution in an alternate form, we are able
to emphasize a number of unusual features: (a) despite the Markovian
nature of the bath, coherence at long times can be retained; (b) the
long--time asymptotic degree of coherence in the system is intertwined
with the asymptotic difference in level populations; (c) the
traditional population--relaxation and decoherence times, $T_1$ and
$T_2$, lose their meaning when the system is in the presence of an
external field, and are replaced by  more general overall timescales;
(d) increasing the field strength, quantified by the Rabi frequency,
$\Omega$, increases the rate of decoherence rather than reducing it, as
one might expect; and (e) maximum asymptotic coherence is reached when
the system parameters satisfy $\Omega^2 = 1/(T_1T_2)$.
\end{abstract}



\maketitle


\section{\label{sec1} Introduction}

A quantum system whose dynamics is of interest is often part of, or
coupled to, a second system whose dynamics is irrelevant.
Examples include the internal quantum dynamics of a molecule in
solution, a qubit imbedded in a solid, the translational motion of a
single particle in a gas, the dynamics of one part of a molecule, etc.
The overall dynamics of the total system is given by the Hamiltonian
\begin{equation}
 \hat{H} = \hat{H}_S + \hat{H}_B + \hat{V}_{SB} ,
 \label{eq:1}
\end{equation}
where $\hat{H}_S$ and $\hat{H}_B$ describe, respectively, the free
evolution of the system of interest (that we henceforth refer to
as the ``system'') and the extraneous degrees of
freedom (termed the ``bath''), and $\hat{V}_{SB}$ accounts for their
interaction.
Ideally, the system dynamics is obtained by solving the Schr\"odinger
equation for the full Hamiltonian, $\hat{H}$, and then averaging over
the bath degrees of freedom.
However, more often than not, this route is intractable.
Hence, it is common to replace the full dynamics associated with
Eq.~(\ref{eq:1}) by an approximate master equation
\cite{percival,breuer,accardi} for the system density matrix,
$\hat{\rho}_S \equiv {\rm Tr}_B \left[ \ \! \hat{\rho} \ \! \right]$.
This type of equation provides a description of the two effects
induced by the bath on the system: population changes and coherence
loss.

Recently, the degree of quantum coherence of a system has become
increasingly important.
For example, both the coherent control of molecular
processes \cite{riceb,brumerb} and quantum manipulations
in quantum computing, quantum information and quantum
cryptography \cite{unruh,ekert,viola,qirefs}, rely upon the ability
to keep the coherence of a system as well as to counter the
decohering effects induced by the environment.
One long--standing approach to reintroduce coherence in a system that
is interacting with a bath is to irradiate the system with a coherent
electromagnetic field.

In this paper we provide new insights into the coherence and
decoherence in a paradigmatic two--level system interacting with a
decohering environment and a resonant continuous--wave (CW)
electromagnetic field.
The model that we focus upon is the standard Bloch equation \cite{bloch}
wherein the bath is Markovian, i.e., the coherence that is transferred
to the bath is lost from the system forever.
Using an alternate to the standard solution to this analytic
problem \cite{torrey,allen}, we show that useful new insights emerge
into the way in which the thermal bath and the external electromagnetic
field
interact to produce and sustain coherence in the system.
Specifically, we emphasize that (a) despite the Markovian nature of
the bath, coherence at long times can be retained; (b) the long--time
asymptotic degree of coherence in the system is intertwined with the
asymptotic difference in level populations; (c) the traditional
population--relaxation and decoherence times, $T_1$ and $T_2$, lose
their meaning when the system is in the presence of an external field,
and are replaced by a more general overall timescale; (d) increasing
the field strength, quantified by the Rabi frequency, $\Omega$,
increases the rate of decoherence rather than reducing it, as one might
expect; and (e) maximum asymptotic coherence is reached when the system
parameters satisfy $\Omega^2 = 1/(T_1T_2)$.

The organization of this work is as follows.
A brief description of the two--level
system is presented in Sec.~\ref{sec2}.
The asymptotic and time--dependent solutions to this model are
discussed in Sec.~\ref{sec3}.
Finally, the main conclusions derived from this work are summarized
in Sec.~\ref{sec4}.



\section{\label{sec2} The two level system}

The dynamical evolution of a two--level system influenced by a thermal
bath can be modelled by means of the master equation
\begin{equation}
 \frac{d\hat{\rho}_S}{dt} =
  - \frac{i}{\hbar} \left[ \hat{H}'(t), \hat{\rho}_S \right]
  - \mathcal{R} \ \! \hat{\rho}_S ,
 \label{eq:3}
\end{equation}
where $\hat{H}'(t) = \hat{H}_S + \hat{H}_{\rm int}(t)$, and
$\mathcal{R}$ is a superoperator describing the evolution of the
bath and its effects on the system (i.e., $\hat{H}_B + \hat{H}_{SB}$).
The evolution of the isolated (free) two--level system is determined
by the Hamiltonian
\begin{displaymath}
 \hat{H}_S = \sum_{i = 1,2} E_i |i\rangle\langle i| ,
\end{displaymath}
and
\begin{displaymath}
 \hat{H}_{\rm int} = - E(t) \sum_{\substack{i,j = 1,2\\i \ne j}}
  \hat{\bf d} |i\rangle\langle j| ,
\end{displaymath}
which accounts for the atom--field interaction within the dipole
approximation, with $E(t) = \mathcal{E} \cos(\omega t + \varphi)$
and $\mathcal{E}$ being the strength of the electromagnetic field.

The simplest model for dynamics of this type is the Markovian Bloch
equation,
\begin{equation}
 \frac{d\rho_{i,j}}{dt} =
  - \frac{i}{\hbar} \left[ \hat{H}'(t), \hat{\rho} \right]_{i,j}
  - \frac{1}{T_{i,j}} \ \! \rho_{i,j} .
 \label{eq:4}
\end{equation}
where $\mathcal{R}$ is written in terms of the phenomenological
relaxation times: $T_1 = T_{i,i}$ and $T_2 = T_{i,j}$ ($i \ne j$).
In the absence of the electromagnetic field, $T_1$ provides the
timescale for changes in the system (eigenstate) populations,
$\rho_{i,i}(t)$, with a rate given by $\Gamma_1 = 1/T_1$.
Due to system--bath elastic collisions, there are random changes in
the system phases that affect the off--diagonal terms $\rho_{i,j}(t)$
($i \neq j$) and subsequently lead to system decoherence at a rate
$\Gamma_2 = 1/T_2$.

In the most general approach the values of $T_1$ and $T_2$ are not
constrained.
For example, Skinner and coworkers have shown \cite{skinner1,skinner2},
using a non--Markovian model of a two--level system linearly and
off--diagonally coupled to a harmonic quantum--mechanical bath, that
$T_2$ can actually be greater than $2 T_1$, with $T_2 = 2 T_1$ in the
weak coupling limit.
In the case of the standard Bloch equation \cite{bloch}, however,
one has
\begin{equation}
 2T_1 \ge T_2
 \label{eq:cond}
\end{equation}
in order to ensure that the reduced dynamics for the system always
leads to completely positive maps of the density matrix \cite{Daffer},
i.e., that ${\rm Tr} \left[ \ \! \hat{\rho}_S^2 \ \! \right] \le 1$
at any time.
Thus, although the Bloch equation is mathematically
well--defined for any values of $T_1$ and $T_2$, they lose their
physical meaning when Eq.~(\ref{eq:cond}) is not satisfied.
Here, the condition (\ref{eq:cond}) is retained throughout the study.

Equation~(\ref{eq:4}) can be solved within the rotating--wave
approximation \cite{allen2} by introducing the following change of
variables \cite{note1}:
\setlength\arraycolsep{2pt}
\begin{eqnarray}
 R_1 & = & 2 \ \! {\rm Im} \left[ \ \! \rho_{12} \ \! \right]
  = -i(\rho_{12} - \rho_{21}) ,
 \nonumber \\
 R_2 & = & 2 \ \! {\rm Re} \left[ \ \! \rho_{12} \ \! \right]
  = \rho_{12} + \rho_{21} ,
 \label{transform} \\
 R_3 & = & \rho_{11} - \rho_{22} .
 \nonumber
\end{eqnarray}
Here, $R_3$ is the difference in population between the two levels,
and $R_1$ and $R_2$ are the imaginary and real components of the
off--diagonal density matrix elements.
With these new variables, Eq.~(\ref{eq:3}) can then be rewritten
\cite{note2} in the standard form of the optical Bloch equations for
a two--level system as
\begin{subequations}
 \setlength\arraycolsep{2pt}
 \begin{eqnarray}
  \frac{d R_1}{dt} & = & - \Gamma_2 R_1 + \Delta R_2 + \Omega R_3 ,
  \label{eq:5a} \\
  \frac{d R_2}{dt} & = & - \Delta R_1 - \Gamma_2 R_2 ,
  \label{eq:5b} \\
  \frac{d R_3}{dt} & = & - \Omega R_1 - \Gamma_1 (R_3 - \tilde{R}_3) ,
  \label{eq:5c}
 \end{eqnarray}
 \label{eq:5}
\end{subequations}
where $\Omega = 2 d_{12} \mathcal{E}/\hbar$ (with $d_{12} = \langle 1 |
\hat{\bf d} | 2 \rangle$) is the Rabi frequency with which the system
oscillates between the two levels in the absence of a bath, and
$\Delta = (E_2 - E_1)/\hbar - \omega$ is the detuning of the laser
frequency $\omega$ from the $|E_1\rangle \to |E_2\rangle$ transition.

The quantity $\tilde{R}_3$ in Eq.~(\ref{eq:5c}) is the thermal
equilibrium population difference to which $R_3$ asymptotically relaxes
in the absence of the external field, and is defined as
\begin{equation}
 \tilde{R}_3 = \frac{1 - e^{- \hbar \omega_{2,1}/k_BT}}
                    {1 + e^{- \hbar \omega_{2,1}/k_BT}} ,
 \label{eq:7}
\end{equation}
with $\omega_{2,1} = (E_2 - E_1)/\hbar$.
$\tilde{R}_3$ indicates the degree of mixedness of the reduced density
matrix at temperature $T$ and in the absence of the external field
($\Omega = 0$).
Two limits are therefore evident: (a) $\tilde{R}_3 \to 1$ if
$T \to 0$, and (b) $\tilde{R}_3 \to 0$ if $T \to \infty$.


\section{\label{sec3} Results and Discussion}


\subsection{\label{sec3A} Asymptotic Coherence}

Our interest focuses on the nature of the system coherence and its
dependence on the system and field parameters.
Two quantities serve as useful measures of coherence:
the purity, $\chi$, and the interference contribution, $\zeta$.
The purity is given by
\begin{eqnarray}
 \chi = {\rm Tr} \left[ \ \! \hat{\rho}_S^2 \ \! \right] & = &
  \rho_{11}^2 + \rho_{22}^2 + 2 |\rho_{12}|^2 \nonumber \\
   & = & \frac{1}{2} + \frac{1}{2} \left(R_1^2 + R_2^2 + R_3^2\right) ,
 \label{eq:9}
\end{eqnarray}
and depends on both level populations and interference.
By contrast, the interference contribution, which we define as
\begin{equation}
 \zeta = |\rho_{12}|^2 = \frac{1}{4} \left(R_1^2 + R_2^2\right) ,
 \label{eq:10}
\end{equation}
describes the coherence in the energy basis.
Hence, though far from an ideal measure of decoherence, $\chi$ is
particularly useful due to its basis--independence.

At long times, when the system reaches the equilibrium, all time
derivatives in Eqs.~(\ref{eq:5}) are zero, and the (asymptotic) value of the
$R_i$ becomes, for the on-resonance case ($\Delta=0$) emphasized below,
\setlength\arraycolsep{2pt}
\begin{eqnarray}
 R_1^{\rm eq} & = & - \frac{\Gamma_1 \Omega}
  {\Gamma_1 \Gamma_2 + \Omega^2} \ \! \tilde{R}_3 ,
 \nonumber \\
 R_2^{\rm eq} & = & 0 ,
 \nonumber \\
 R_3^{\rm eq} & = & \frac{\Gamma_1 \Gamma_2}
  {\Gamma_1 \Gamma_2 + \Omega^2} \ \! \tilde{R}_3 .
 \nonumber
\end{eqnarray}
Consequently, inverting Eqs.~(\ref{transform}) gives the reduced
density matrix elements \cite{note3}:
\begin{subequations}
 \setlength\arraycolsep{2pt}
 \begin{eqnarray}
  \rho_{12}^{\rm eq} & = & \left( \rho_{21}^{\rm eq}\right)^*
   = \frac{1}{2} \bigg[ \frac{T_2 \Omega}{1 + T_1 T_2\Omega^2}
    \bigg] e^{i(2\varphi - \pi/2)} \tilde{R}_3 ,
  \label{eq:8a} \\
  \rho_{11}^{\rm eq} & = & \frac{1}{2} \bigg[ 1 + \frac{\tilde{R}_3}
    {1 + T_1 T_2 \Omega^2} \bigg] ,
  \label{eq:8b} \\
  \rho_{22}^{\rm eq} & = & \frac{1}{2} \bigg[ 1 - \frac{\tilde{R}_3}
    {1 + T_1 T_2 \Omega^2} \bigg] ,
  \label{eq:8c}
 \end{eqnarray}
 \label{eq:8}
\end{subequations}
where it is apparent that $\rho_{11}^{\rm eq} \geq \rho_{22}^{\rm eq}$.
Observe that
\begin{equation}
 \rho_{12}^{\rm eq} =
  -\frac{1}{2} \ \! T_2 \Omega \ \! e^{i(2\varphi + \pi/2)}
   \left[ \rho_{11}^{\rm eq} - \rho_{22}^{\rm eq} \right] ,
  \label{eq:8bis}
\end{equation}
giving a relationship between the coherence elements of the
reduced density matrix and the population difference.

Numerous features regarding the coherence at long times are evident
from Eqs.~(\ref{eq:8}).
Obviously, the coherence is totally lost ($\zeta=0$) in the absence of
the field ($\Omega=0$).
Most significantly, as emphasized further below, the asymptotic extent
of the coherence (as manifest in the value of $\rho_{12}^{\rm eq}$) is
directly proportional to $\tilde{R}_3$, the difference between the
final populations of each state.
Thus, for example, if the temperature is high, $\tilde{R}_3 =0$
and $\zeta=0$ regardless of the initial conditions.
Similarly, at lower temperatures, and also regardless of initial
conditions, the asymptotic value of $\zeta$ can be nonzero, and the
system can have therefore gained coherence due to the combined
influence of the non--zero asymptotic population difference (related
to the nature of the environment) and the field.

Studies that focus solely on decoherence \cite{viola} often neglect
population--relaxation processes by setting $T_1 = \infty$.
In this case [from Eq.~(\ref{eq:8a})], all coherence is lost at
equilibrium, with $\rho_{12}^{\rm eq} = 0$.
Hence, long--time coherences that can exist in the case of finite
$T_1$ are missed.
This reliance of long--time coherence emphasizes the significant
interplay between the long--time population difference, $R_3^{\rm eq}$,
the time that it takes to reach that limit (as manifest in $T_1$
timescales), and the long--time coherence, $\rho_{12}^{\rm eq}$.

Interestingly, coherence loss also occurs with large $\Omega$, as
is evident in Eq.~(\ref{eq:8a}).
That is, $|\rho_{12}^{\rm eq}|$ is an increasing function of $\Omega$
until $\Omega^2 = 1/(T_1T_2)$, at which point $|\rho_{12}^{\rm eq}| =
\sqrt{T_2/T_1} \ \! \tilde{R}_3/4$.
Increasing $\Omega$ beyond this value causes a decrease in the
asymptotic coherence.
Note that in the particular case where $T_1$ is assumed very large,
the asymptotic coherence is a decreasing function of $\Omega$ for
almost all $\Omega$ values.


\subsection{\label{sec3B} Time Evolution}

The analytic solution to Eqs.~(\ref{eq:5}) for the time evolution of
the $R_i$ is known \cite{torrey}.
Here we rewrite it in a somewhat more enlightening form.

Consider the case of on--resonance excitation ($\Delta = 0$).
Laplace transforming Eq.~(\ref{eq:5c}), and defining the transforms
with over--bars, leads to the following linear system:
\begin{subequations}
 \setlength\arraycolsep{2pt}
 \begin{eqnarray}
  (s + \Gamma_2) \bar{R}_1 + \Omega \bar{R}_3 & = & R_1^{(0)} ,
  \label{eq:apc2a} \\
  (s + \Gamma_2) \bar{R}_2 & = & R_2^{(0)} ,
  \label{eq:apc2b} \\
  (s + \Gamma_1) \bar{R}_3 - \Omega \bar{R}_1 & = & R_3^{(0)}
    + \Gamma_1 \tilde{R}_3/s ,
  \label{eq:apc2c}
 \end{eqnarray}
 \label{eq:apc2}
\end{subequations}
where the $R_i^{(0)}$ are the initial conditions for each of the
variables.

As seen from Eq.~(\ref{eq:apc2b}), the time--dependence of $R_2$,
\begin{equation}
  R_2(t) = R_2^{(0)} e^{- \Gamma_2 t} ,
 \label{eq:apc3}
\end{equation}
is readily obtained, since this variable is uncoupled from the
other two.
Therefore, for fixed $T_2$ and $T_1 > T_2$, the dynamics is described
by $R_1$ and $R_3$ beyond times on the order of $T_2$.

To obtain $R_1$ and $R_3$, the value of $\bar{R}_1$ resulting from
Eq.~(\ref{eq:apc2c}) is substituted into (\ref{eq:apc2a}), giving
%
\begin{equation}
 \left[(s + \Gamma_1)(s + \Gamma_2) + \Omega^2\right] \bar{R}_3 =
  \Omega R_1^{(0)} + (s + \Gamma_2) R_3^{(0)}
   + \frac{(s + \Gamma_2) \Gamma_1}{s} \ \! \tilde{R}_3 ,
 \label{eq:apc4}
\end{equation}
where the roots of the factor multiplying $\bar{R}_3$ are
$s_\pm = - \alpha \pm \beta$, with
\begin{displaymath}
 \alpha = \frac{\Gamma_1 + \Gamma_2}{2} , \ \ \
 \beta = \frac{\sqrt{(\Gamma_2 - \Gamma_1)^2 - 4 \Omega^2}}{2} .
\end{displaymath}
Depending on the magnitude of the discriminant of $\beta$, three cases
result.
If $|\Gamma_2 - \Gamma_1| \neq 2 \Omega$, then
\begin{equation}
 \bar{R}_3 = \frac{D_3}{s} + \frac{B_3}{s - s_-} + \frac{C_3}{s - s_+} ,
 \label{eq:apc6}
\end{equation}
with
\begin{subequations}
 \setlength\arraycolsep{2pt}
 \begin{eqnarray}
  D_3 & = & \frac{\Gamma_1 \Gamma_2}{s_{-}s_{+}} \tilde{R}_3
   = \frac{\Gamma_1 \Gamma_2}{\Gamma_1 \Gamma_2 + \Omega^2}
    \tilde{R}_3 ,
  \label{eq:apc7a} \\
  B_3 & = & \frac{s_{-}^2 R_3^{(0)} + s_{-} \Lambda
   + \Gamma_1 \Gamma_2 \tilde{R}_3}{s_{-} (s_{-} - s_{+})} ,
  \label{eq:apc7b} \\
  C_3 & = & \frac{s_{+}^2 R_3^{(0)} + s_{+} \Lambda
   + \Gamma_1 \Gamma_2 \tilde{R}_3}{s_{+} (s_{+} - s_{-})} ,
  \label{eq:apc7c}
 \end{eqnarray}
 \label{eq:apc7}
\end{subequations}
where $\Lambda \equiv \Omega R_1^{(0)} + \Gamma_2 R_3^{(0)} +
\Gamma_1 \tilde{R}_3$.
The inverse Laplace transform of Eq.~(\ref{eq:apc6}) leads to
\begin{equation}
 R_3(t) = D_3 + \left(B_3 e^{-\beta t} + C_3 e^{\beta t}\right)
  e^{-\alpha t} .
 \label{eq:apc8}
\end{equation}
Now, introducing Eq.~(\ref{eq:apc8}) into Eq.~(\ref{eq:apc2a}),
we obtain for the imaginary part of the coherence
\begin{equation}
 R_1(t) = D_1 + \left(B_1 e^{-\beta t} + C_1 e^{\beta t}\right)
  e^{-\alpha t} ,
 \label{eq:apc9}
\end{equation}
with
\setlength\arraycolsep{2pt}
\begin{eqnarray}
 D_1 & = & - \frac{\Gamma_1 \Omega}{\Gamma_1 \Gamma_2 + \Omega^2}
  \tilde{R}_3 , \nonumber \\
 B_1 & = & \frac{(s_{-} + \Gamma_1)}{\Omega} \ \! B_3 , \nonumber \\
 C_1 & = & \frac{(s_{+} + \Gamma_1)}{\Omega} \ \! C_3 . \nonumber
\end{eqnarray}
Therefore, for $|\Gamma_2 - \Gamma_1| \neq 2 \Omega$, both $R_1$ and
$R_2$ decay exponentially.
However, for $|\Gamma_2 - \Gamma_1| < 2 \Omega$, $\beta$ is imaginary,
and $R_1$ and $R_3$ oscillate as they decay.
Below, the time--dependent factors depending on $\beta$ and that are
responsible for the fine structure of the decays will be labelled $F_i$.

Specific results for the case of ``critical damping'',
$|\Gamma_2 - \Gamma_1| = 2 \Omega$, can be obtained by considering
the limit of the previous expressions when $\beta \to 0$ (for which
$s_\pm = s_0 = - \alpha$).
In this case both $R_1$ and $R_3$ exponentially decay with rate
$\alpha$, although the $F_i$ are linear functions of time.

These results [Eqs.~(\ref{eq:apc8}) and (\ref{eq:apc9})] allow direct
consideration of the rates of falloff observed for the populations and
coherences in the presence of a non--zero field.
Specifically, if the field is strong ($2\Omega \ge |\Gamma_2
- \Gamma_1|$) both the populations and the coherences decay with an
overall rate $\alpha \le \Gamma_2$ whenever $T_1 \ge T_2$, albeit
with superposed oscillations.
Thus, in principle, there is no distinction between the falloff rates
of the diagonal and off--diagonal elements of the density matrix.
This is also evident in $\chi$, where the long--time decay
goes as $2\alpha = \Gamma_1 + \Gamma_2$, and it is not possible to
clearly distinguish between two different timescales in its evolution.
Nonetheless, since the $B_i$ and $C_i$ depend on $\Omega$, the
amplitude of the $F_i$ can be greater than unity, and therefore may
lead to slower $\chi$ decay rates than $2\alpha$ at early times despite
being multiplied by the exponential.
In the limit of $\Omega \to \infty$, the $F_i$ become highly
oscillatory functions of time with decreasing amplitude about
$F_i = 1$, and the decay rate exactly corresponds to $2\alpha$.
This can be easily seen by substituting Eqs.~(\ref{eq:apc3}),
(\ref{eq:apc8}), and (\ref{eq:apc9}) into Eq.~(\ref{eq:9}).
For example, for an initial state comprised of population in the state
$|1\rangle$ (i.e., $R_1^{(0)} = R_2^{(0)} = 0$, $R_3^{(0)} = 1$), one
obtains
%
\begin{equation}
 \chi = \frac{1}{2} + \frac{1}{2} \left[ 1
  + 2 \left(\frac{\Gamma_2 - \Gamma_1}{2\beta}\right)^2
   \sin^2 \beta t + \left(\frac{\Gamma_2 - \Gamma_1}{2\beta}\right)
   \sin 2\beta t \right] e^{-2\alpha t} ,
 \label{eq:x}
\end{equation}
which effectively approaches
\begin{equation}
 \chi = \frac{1}{2} + \frac{1}{2} \ \! e^{-2\alpha t} ,
 \label{eq:xx}
\end{equation}
when $\Omega \to \infty$ (and therefore $\beta \to \Omega$).

If the field is weaker ($2\Omega < |\Gamma_2 - \Gamma_1|$) there are
two system decay rates for $R_1$ and $R_3$, $\alpha \pm \beta$, with
the smaller of the two dominating at longer time.
In the parameter region where $2 \Omega < |\Gamma_2 - \Gamma_1|$, the
decay rate increases with increasing $\Omega$, evidently reaching its
maximum, in this region, at critical damping.
However, numerical evidence reported below shows that the decay rate
continues its increase with increasing $\Omega$ in the stronger field
region, where $2 \Omega \ge |\Gamma_2 - \Gamma_1|$.
Note that only in the limit of very small $\Omega$ does the standard
field--free interpretation ($R_1$ and $R_2$ decaying with rate
$\Gamma_2$, and $R_3$ with rate $\Gamma_1$) apply, since both $B_3$
and $C_1$ approach zero in this limit.


\subsubsection{$T = \infty$}

\begin{figure}[!ht]
 \vspace{2cm}
 \includegraphics[width=7cm]{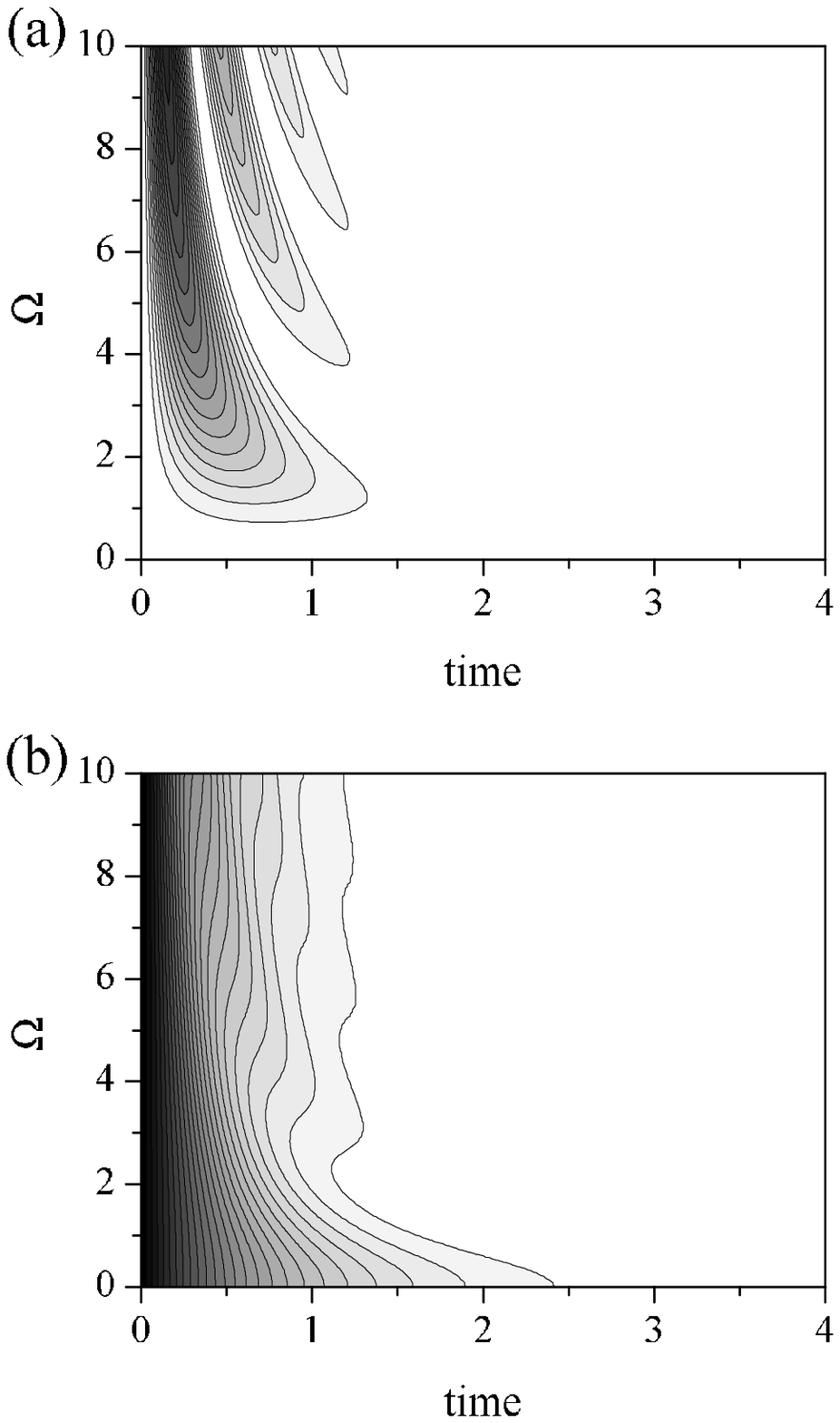}
 \caption{\label{fig:1}
  (a) $\zeta$ and (b) $\chi$ as a function of the Rabi frequency and
  time for $T_1 = 1.5$, $T_2 = 0.5$, and $\tilde{R}_3 = 0$.
  For all figures the darkest shading corresponds to 0.20 for
  $\zeta$--plots, and 20 increasing lighter shadings are used until
  $\zeta = 0$.
  For $\chi$--plots the darkest shading corresponds to 1 and 20
  increasing lighter shadings are used until $\chi = 0.5$.}
\end{figure}

Consider first the case where $T=\infty$.
Here, the field is unable to beat out, at long time, the
thermalizing effects of the bath, thus leading to zero asymptotic
coherence.
This happens even if the field is CW and is on for all time.
The coherence dynamics on the way to the asymptotic result is of
interest.
As an indication of the non--intuitive nature of the associated
relaxation timescales, consider the results for $\zeta$ and $\chi$,
shown in Figs.~\ref{fig:1}(a) and (b).
A grey scale is used to indicate the magnitude of $\zeta$ and $\chi$
as a function of $t$ and $\Omega$ for an initial state comprised of
population in $|1\rangle$.
A number of interesting features can be observed.
First, as noted above, the decay rate increases with increasing $\Omega$,
even beyond the case of $2\Omega = |\Gamma_2 - \Gamma_1| \simeq 1.33$.
Further, the onset of oscillatory falloff above this value is clearly
visible.
The (overall) falloff timescale, given by $(2\alpha)^{-1}$, is in
this case $\approx 0.375$, making evident the inseparability of the
meaning of $T_1$ and $T_2$ (i.e., one cannot observe two clearly
separated decay timescales in $\chi$).

\begin{figure}[!ht]
 \vspace{3.5cm}
 \includegraphics[width=7.3cm]{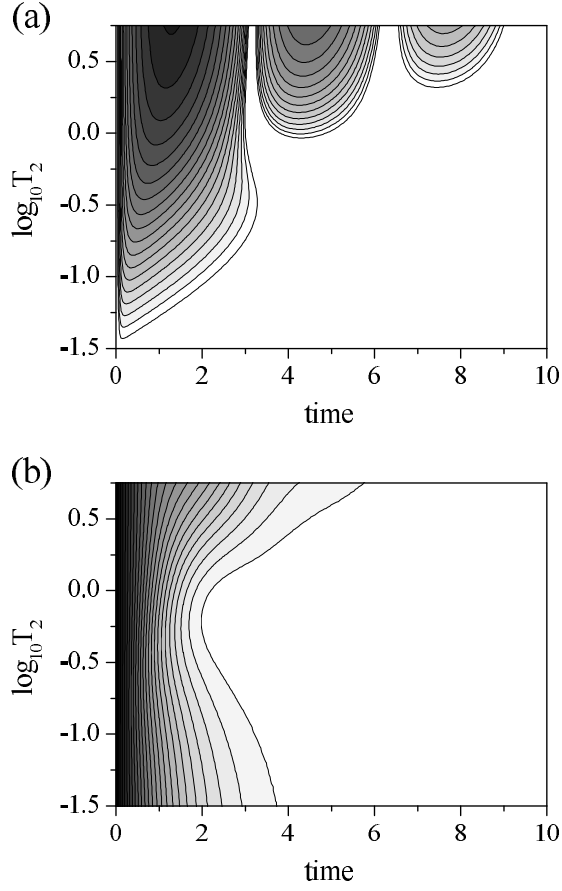}
 \caption{\label{fig:2}
  (a) ${\rm Log}_{10} \ \! \zeta$ and (b) $\chi$ as a function
  of $\log_{10} T_2$ and time for $T_1 = 2.5$, and
  $\tilde{R}_3 = 0$.
  The upper limit of the ordinate scale corresponds to $T_2 = T_1/2$,
  as in Eq.~(\ref{eq:cond}).
  For all the $\log_{10} \ \! \zeta$--plots the darkest shading
  corresponds to $\zeta = 0.30$, and 20 increasing lighter shadings
  are used until $\zeta = 0.0003$.}
\end{figure}

\begin{figure}[!ht]
 \vspace{7cm}
 \includegraphics[width=7cm]{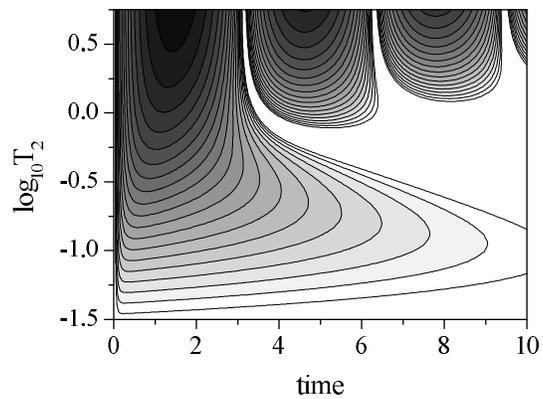}
 \caption{\label{fig:3}
  ${\rm Log}_{10} \ \! \zeta$ as a function of $\log_{10} T_2$ and
  time for $T_1 = 10^4$, and $\tilde{R}_3 = 0$.
  The upper limit on the ordinate is as in Fig.~\ref{fig:2}.}
\end{figure}

The alternative perspective in Figs.~\ref{fig:2}(a) and (b), where
$\zeta$ and $\chi$ are shown as a function of $\log_{10} T_2$ and time
for fixed $T_1=2.5$ and $\Omega=1$ is also enlightening.
In this case, the rate of decay of $\zeta$ and $\chi$ are seen to fall
off increasingly rapidly as $T_2$ becomes {\it larger}, until
$|\Gamma_2 - \Gamma_1| = 2\Omega$, (here corresponding to log$_{10}T_2 = -0.2$).
At this point, with increasing $T_2$, the rate of falloff slows down
and oscillates (these oscillations are better appreciated in the case
of $\zeta$).
The effect becomes increasingly exaggerated with larger $T_1$, as is
evident from Fig.~\ref{fig:3}, that shows log $\zeta$ for $T_1 = 10^4$.
Clearly, regarding $T_2$ as the timescale for loss of coherence makes
little sense in this context.

\begin{figure}[!ht]
 \vspace{.5cm}
 \includegraphics[width=7cm]{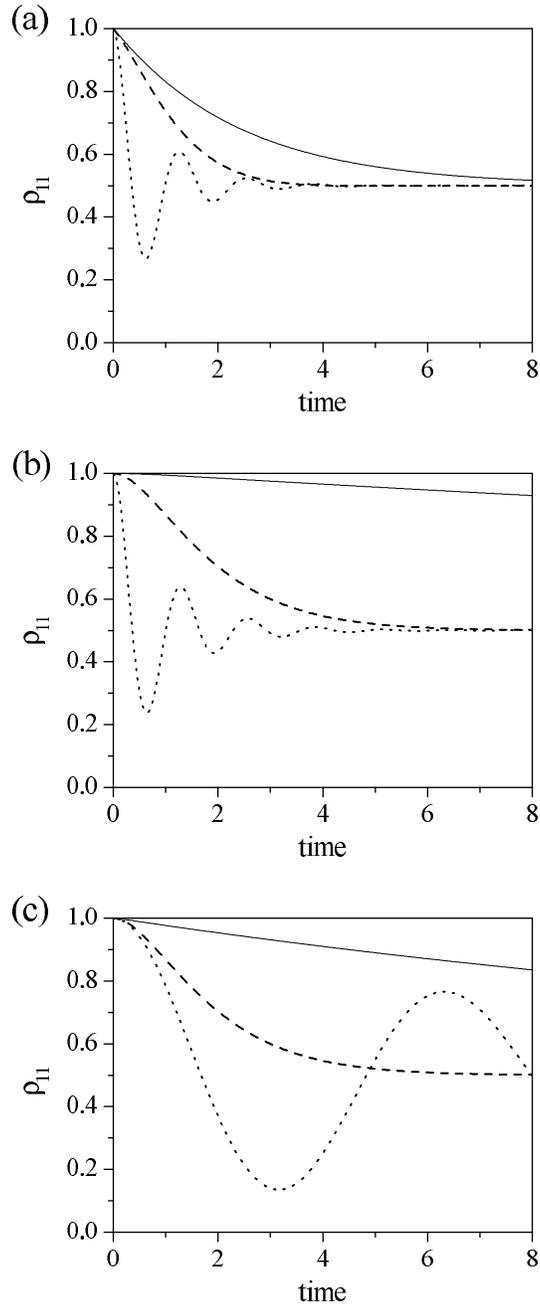}
 \caption{\label{fig:4}
  $\rho_{11}(t)$ for three different values of the Rabi frequency:
  $\Omega = 0.2$ (solid line), $\Omega = 1$ (dashed line), and
  $\Omega = 5$ (dotted line).
  To compare, two different values of $T_1$ are considered (with
  $T_2 = 0.5$): (a) $T_1 = 2.5$ and (b) $T_1 = 10^4$.
  In (c), $\rho_{11}(t)$ for three different values of $T_2$ (with
  $T_1 = 10^4$ and $\Omega = 1$): $T_2 = 0.05$ (solid line),
  $T_2 = 0.5$ (dashed line), and $T_2 = 5$ (dotted line).}
\end{figure}

Related insights, here into the reinterpretation of $T_1$, are provided
by considering, for example, populations $\rho_{11}$, as shown in
Fig.~\ref{fig:4}.
Focusing on Figs.~\ref{fig:4}(a) and (b), we see that in the case where
the field is weak (the solid curve for $\Omega = 0.2$) the rate of
population relaxation does indeed decrease with increasing $T_1$.
However, the value of $T_1$ loses this qualitative meaning when the
field strength is increased, as is evident by comparing the dashed
curves and dotted curves, corresponding to stronger fields, in each of
panels (a) and (b).
That is, despite the fact that $T_1$ differs in these two panels by a
factor of 4000, the populations relax at qualitatively similar rates.
Further, note that the increasing field strength tends to be associated
with an increase in rate of the population relaxation.
Similarly, Fig.~\ref{fig:4}(c) shows that the rate of population
relaxation depends on $T_2$ as well as $T_1$.
Notice that here, effectively, the fastest relaxation is achieved for
$\Gamma_2 \simeq 2\Omega$, i.e., $|\Gamma_2 - \Gamma_1| = 2 \Omega$
[as one can also see in Fig.~\ref{fig:2}(b)].
Clearly, the presence of the field imposes less meaning to the
traditional values of $T_1$ and $T_2$.


\subsubsection{Finite $T$}

\begin{figure}[!ht]
 \vspace{3.5cm}
 \includegraphics[width=7cm]{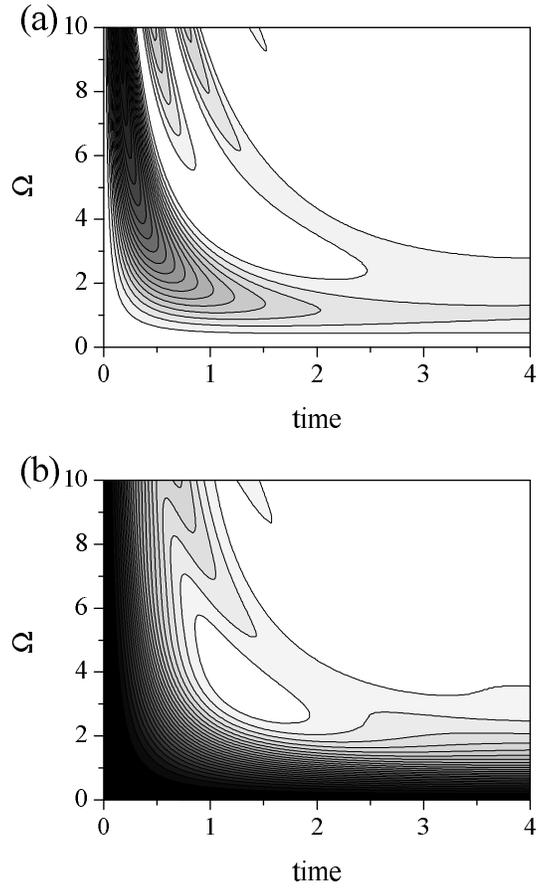}
 \caption{\label{fig:5}
  (a) $\zeta$ and (b) $\chi$ as a function of the Rabi frequency
  and time for $T_1 = 1.5$, $T_2 = 0.5$, and $\tilde{R}_3 = 1$.}
\end{figure}

Additional significant effects arise when the temperature is not
infinite and the asymptotic population difference, $\tilde{R}_3$,
is nonzero.
Consider, for example, the extreme case were $\tilde{R}_3 =1$, shown
in Figs.~\ref{fig:5}(a) and (b), for $\zeta$ and $\chi$, respectively.
Specifically, a comparison with Figs.~\ref{fig:1}(a) and (b) shows
large qualitative differences at all but the highest $\Omega$.
In particular, two effects are evident.
First, the nonzero asymptotic value of $\zeta$, due to the nonzero
$\tilde{R}_3$, is clear.
Second, significantly, both $\zeta$ and $\chi$ are seen to show regions
of $\Omega$ where the function decays to the incoherent limit (zero in
the case of $\zeta$; one--half in the case of $\chi$), but then
reestablishes coherence to reach a long time value that contains
coherence.
This is clear, for example, in the case of $\zeta$ for $\Omega$
between approximately 2.1 and 5.4.
Similar results are seen for $\chi$ for a range of $\Omega$
between 2.4 and 5.7.
Hence, it would be misleading to suggest that decay to the incoherent limit
implies no reestablishment of coherence.
For example, for $\Omega = 4$, the system undergoes a coherence
revival for a time interval $\Delta t \approx 0.4$, and for
$\Omega \simeq 2.3$ the revival takes place at $t \approx 2.5$
and the coherence remains permanently.
Both of the above effects weaken with decreasing
$\tilde{R}_3$.

\begin{figure}[!ht]
 \vspace{.5cm}
 \includegraphics[width=7cm]{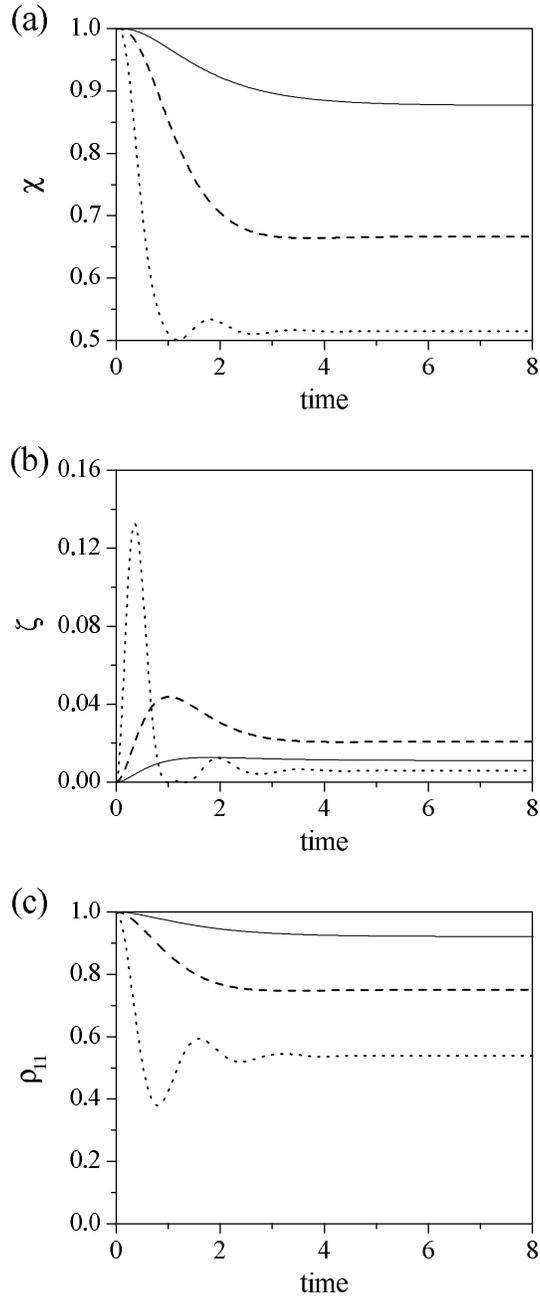}
 \caption{\label{fig:6}
  (a) $\chi$, (b) $\zeta$, and (c) $\rho_{11}$ for three different
  values of the Rabi frequency: $\Omega = 0.5$ (solid line),
  $\Omega = \Omega_r=1.15$ (dashed line), and $\Omega = 4$ (dotted line).
  In all cases: $\tilde{R}_3 = 1$, $T_1 = 1.5$, $T_2 = 0.5$.}
\end{figure}

Finally, Fig.~\ref{fig:6} stresses the fact that the maximum amount
of asymptotic coherence depends on the three parameters defining the
system evolution. Thus, given $T_1$ and $T_2$, the value of the
frequency for which one has maximum asymptotic coherence is
$\Omega_r = 1/\sqrt{T_1T_2}$ [see Fig.~\ref{fig:6}(b)], which leads
to $|\rho_{12}^{\rm eq}| = \sqrt{\frac{T_2}{T_1}} \tilde{R}_3/4$. In
particular, for the values used in Fig.~\ref{fig:6}, $\Omega_r
\simeq 1.15$.


\section{\label{sec4} Summary}

Aspects of the ubiquitous Optical Bloch equations have been
examined with a particular focus on the coherence of the system.
The qualitative view that populations relax with rates of $1/T_1$, and
coherences relax with rates $1/T_2$ is seen to be misleading when the
system is irradiated by an external field with Rabi frequency $\Omega$.
Similarly, increasing $\Omega$ is shown to increase the rate of
decoherence, contrary to simple intuition. Finally, the possibility
that the coherence can recur to nonzero asymptotic values after having
decayed to zero is noted.


\begin{acknowledgments}
This work was supported by the Natural Sciences and Engineering
Research Council of Canada (NSERC).
\end{acknowledgments}



\newpage
\section*{}

\begin{thebibliography}
\bibitem{note1}
 We follow here the same notation as in Ref.~\onlinecite{brumerb}.
\bibitem{note2}
 Note that, on resonance ($\Delta = 0$), as considered here, the
 equations of motion for $R_1$ and $R_3$ can be expressed as
 %
 \setlength\arraycolsep{2pt}
 \begin{eqnarray}
  \ddot{R}_1 + (\Gamma_1 + \Gamma_2) \dot{R}_1
   + (\Omega^2 + \Gamma_1\Gamma_2) R_1 & = &
    - \Gamma_1\Omega \tilde{R}_3 , \nonumber \\
  \ddot{R}_3 + (\Gamma_1 + \Gamma_2) \dot{R}_3
   + (\Omega^2 + \Gamma_1\Gamma_2) R_3 & = &
    \Gamma_1\Gamma_2 \tilde{R}_3 , \nonumber
 \end{eqnarray}
 %
 which are isomorphic to those of a forced damped oscillator, with
 damping constant $\Gamma = \Gamma_1 + \Gamma_2$, natural harmonic
 frequency $\Omega_0^2 = \Omega^2 + \Gamma_1\Gamma_2$, and driving
 forces $\mathcal{F}_1 = - \Gamma_1\Omega \tilde{R}_3$ and
 $\mathcal{F}_2 = \Gamma_1\Gamma_2 \tilde{R}_3$, respectively, for
 each variable.
 This analogy provides considerable insight into the evolution of
 the two--level system.
\bibitem{note3}
 As in Ref.~\onlinecite{brumerb}, $\rho_{12}$ is replace by
 $\rho_{12} \exp (-2i\varphi)$, this explaining the $2\varphi$ phase
 factor in Eq.~(\ref{eq:8a}).
\end{thebibliography}
\end{document}